\documentclass[]{spie}  %>>> use for US letter paper
\usepackage{amsmath,amsfonts,amssymb}
\usepackage{graphicx}
\usepackage[colorlinks=true, allcolors=blue]{hyperref}

\title{Experimental test of a 40\,cm-long R=100\,000 spectrometer for exoplanet characterisation}

\author[a]{Guillaume Bourdarot}
\author[a]{Etienne Le Coarer}
\author[a]{David Mouillet}
\author[a]{Jean-Jacques Correia}
\author[a]{Laurent Jocou}
\author[a]{Patrick Rabou}
\author[a]{Alexis Carlotti}
\author[a]{Xavier Bonfils}
\author[b]{\'Etienne Artigau}
\author[b]{Philippe Vallee}
\author[b]{Rene Doyon}
\author[a]{Thierry Forveille}
\author[a]{Eric Stadler}
\author[a]{Yves Magnard}
\author[c]{Arthur Vigan}

\affil[a]{Universit\'e Grenoble Alpes, CNRS, IPAG, F-38000 Grenoble, France}
\affil[b]{Institut de Recherche sur les Exoplan\`etes (IREx), D\'epartement de Physique, Universit\'e de Montr\'eal, C.P. 6128, Succ. Centre-Ville, Montr\'eal, QC, H3C 3J7, Canada}
\affil[c]{Aix-Marseille University, CNRS, LAM, UMR 7326, 13388 Marseille, France}

\authorinfo{Contact author :\\ \textbf{G.Bourdarot} : E-mail: guillaume.bourdarot@univ-grenoble-alpes.fr, Telephone: (+33)6 72 39 14 98
}

\begin{document} 
\maketitle

\begin{abstract}
High-resolution spectroscopy is a key element for present and future astronomical instrumentation. In particular, coupled to high contrast imagers and coronagraphs, high spectral
resolution enables higher contrast and has been identified as a very powerful combination to
characterise exoplanets, starting from giant planets now, up to Earth-like planet eventually for the
future instruments.
%However, existing high-resolution NIR spectrometer on large AO-corrected
%telescope are not yet optimized for such scenarii, namely single-mode operation and high-efficiency
%in a targeted spectral domain ; the lack of versatility and the bulk associated to such spectrometers
%being also a limiting factor to this purpose.\newline \newline
In this context, we propose the implementation of an innovative echelle spectrometer based on the use of VIPA (Virtually Imaged Phased Array, Shirasaki 1996). The VIPA itself is a particular kind of Fabry-P\'erot interferometer, used as an angular disperser with much greater dispersive power than common diffraction grating.
%Thus, it combines the high-efficiency and high-resolution of
%the Fabry-P\'erot interferometer in an echelle-spectrometer scheme. 
%By increasing the dispersive power compared to a classical diffraction grating, it enables equally to shrink the %size of the whole spectrometer while keeping the same spectral resolution, or to increase the spectral %resolution while keeping the same size ; practically, it is best suited for high ($R>50\:000$), or even very %high ( $R>200\:quad 000$) spectral resolution. 
The VIPA is an efficient, small component (3\,cm $\times$ 2.4\,cm), that takes the very advantage of single mode injection in a versatile design. The overall instrument presented here is a proof-of-concept of a compact, high-resolution ($R>80\;000$) spectrometer, dedicated to the $H$ and $K$ bands, in the context of the project ``High-Dispersion Coronograhy`` developed at IPAG. The optical bench has a foot-print of 40\,cm $\times$ 26\,cm ; it is fed by two Single-Mode Fibers (SMF), one dedicated to the companion, and one to the star and/or to a calibration channel, and is cooled down to 80\,K.

This communication first presents the scientific and instrumental context of the project, and the principal merit of single-mode operations in high-resolution spectrometry. After recalling the physical structure of the VIPA and its implementation in an echelle-spectrometer design, it then details the optical design of the spectrometer. In conclusion, further steps (integration, calibration, coupling with adaptive optics) and possible optimization are briefly presented. 
\end{abstract}

% Include a list of keywords after the abstract 

\section{INTRODUCTION}
\label{sec:intro}  % \label{} allows reference to this section
\subsection{High-dispersion Coronography}
\begin{figure} [ht]
  \begin{center}
  \begin{tabular}{c} %% tabular useful for creating an array of images 
  \includegraphics[height=7.5cm]{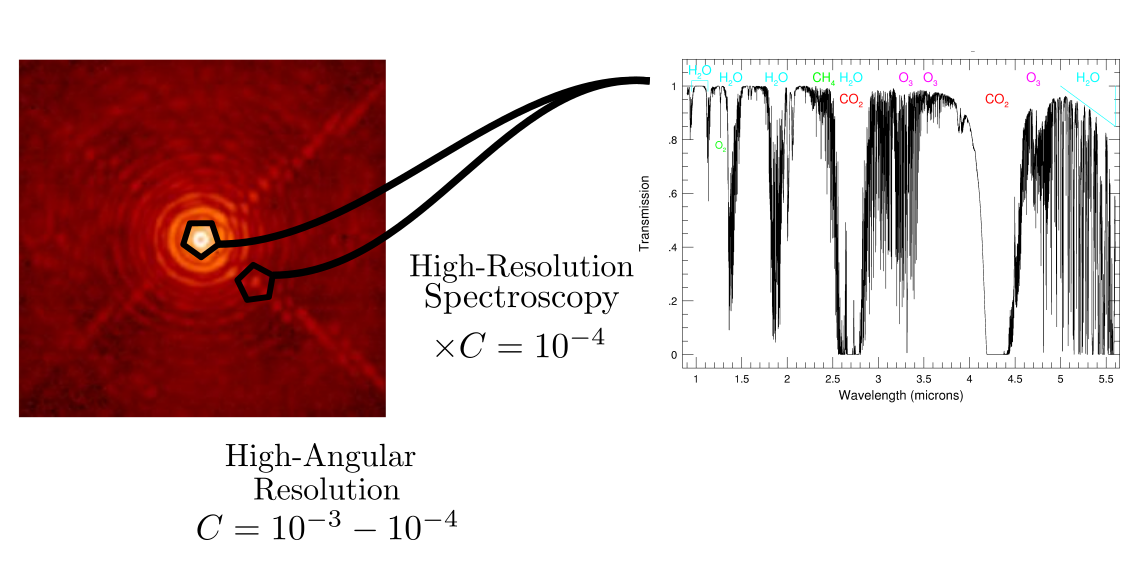}
  \end{tabular}
  \end{center}
  \caption{Principles of High-Dispersion Coronography.}
%>>>> use \label inside caption to get Fig. number with \ref{}
  { \label{fig:HDC} 
}
\end{figure}

Characterisation of exoplanet atmospheres requires to detect extremely low-amplitude signals coming from the planet companion. Transit spectroscopy is for the moment the most successful technique to detect and identify chemical components in exoplanet atmospheres, but still suffer of limited detection contrast and of degeneracies due to the light coming from the host star. On the other hand, high angular resolution gives the capability to resolve spatially the companion, but is still limited by the overwhelming flux ratio between the planet and its host star, as well as the speckle noise floor. A promising avenue for exoplanet characterisation is to cumulate the advantages of these techniques by coupling high-resolution spectroscopy with high resolution imaging (Fig.\ref{fig:HDC}), a technique called "High Dispersion Coronography\cite{Riaud07,Lovis16, Snellen2013, Kasper2016, Wang17, mawet17}. Residual speckles of an AO-corrected image can usually be mis-interpreted as a candidate companion. Spectral information from an IFS provides a significant improvement to discriminate a speckle from a companion, but this remains limited as the spatial and spectral information is partially mixed within the IFS and the a posteriori data reduction. Residual speckles still produce spectral artefacts that may resemble to a low resolution spectrum of a companion. This limitation of speckle noise is solved at high resolution as the companion spectrum (with a different atmosphere composition and/or orbital velocity) differs from the stellar spectrum, regardless of the potential effect of speckles on the continuum.

In this context, the coupling of high-resolution spectrographs with existing extreme AO systems (SPHERE, SCExAO, Keck,...) is a major step to be done. This coupling can already start by using existing spectrographs, such as the CRIRES spectrograph at the VLT. However, it is clear that, if a spectrograph had to be designed for this purpose, it could have a different design. In particular, in a spectrograph design, single-mode operation leads to significant advantages \cite{Jovanovic16,Lexi16} :

\begin{itemize}
\item elementary \textbf{optical \'etendue} $\propto \lambda^2$ : this enables to reach high or even very high spectral resolution, in a simplified optical design, which can even integrate off-the-shelf components;
\item \textbf{compactness} : as the optical \textit{\'etendue} is smaller, the footprint is also decreased, compared with existing instruments with comparable spectral resolution;
\item \textbf{versatility} : the spectrograph is connected to standard optical fibers, and could be transported in order to be coupled to existing AO systems.
\end{itemize}

%We can also note that single-mode design requires less pixels per resolution element compared to multimode design - without pupil slicer -, which results in a greater available spectral range if the total number of pixels is the same.

\subsection{Basic spectrograph equations}
More generally, the fundamental advantage of single-mode operations can be summarized in the fact that, in a spectrometer, a compromised has to be reached between the spectral resolution and the integrated \textit{optical \'etendue}, which could be roughly described in the case of dispersive spectrometer by the well-known luminosity-resolution product\cite{Jacquinot54,Schroeder2000} in  Eq.(\ref{eq:LR}) :

\begin{equation}
\label{eq:LR}
LR=\tau\frac{\pi}{4}D\phi'\lambda A d_2
\end{equation}

with $R=\lambda/d\lambda$ the spectral resolution, $L$ the luminosity of the object, $\phi'$ diameter of seeing disk or angular height of the slit, $A$ the angular dispersion, $d_2$ the diameter of camera lens, $\tau$ the total throughput of the instrument and D the diameter of the telescope. Thus, compared to an existing spectrograph such as CRIRES, decreasing the integrated optical \textit{\'etendue} allows equally : to shrink the size of the spectrometer, the spectral resolution being equal ; or 2) to increase spectral resolution, the size being equal.
 
\subsection{Advantages of VIPA spectrograph}
In this context, the VIPA gives the opportunity to take full advantage of both :
\begin{itemize}
\item single-mode operation;
\item very high angular dispersion (see section ~\ref{sec:dispersion}).
\end{itemize} 
The use of VIPA has already been demonstrated in telecom \cite{Shirasaki96}, a context in which it was invented, and used afterwards for frequency comb applications \cite{Diddams07} to reach very high spectral resolution (of the order of $R=100\,000-1\,000\,000$), and was recently introduced to astrophysics instrumentation \cite{Bourdarot17}.

\section{General concepts of a VIPA echelle spectrometer}
This section details the general approach adopted for the design of an echelle spectrograph based on the use of VIPA. After recalling the physical structure of the VIPA, which is a particular class of Fabry-P\'erot interferometer, the design starts by setting the resolution of the VIPA (which can be very high, precisely given the Fabry-P\'erot structure). Once the resolution is set, the cross dispersion has to be adapted to the Free Spectral Range (FSR) of the VIPA. Finally, as in a diffraction grating, a careful attention has to be paid to the energy distribution inside the different diffraction peaks, which will be related to the notion of Free Angular Range.

\subsection{Physical structure of the VIPA}
\begin{figure} [ht]
  \begin{center}
  \begin{tabular}{c} %% tabular useful for creating an array of images 
  \includegraphics[height=8.5cm]{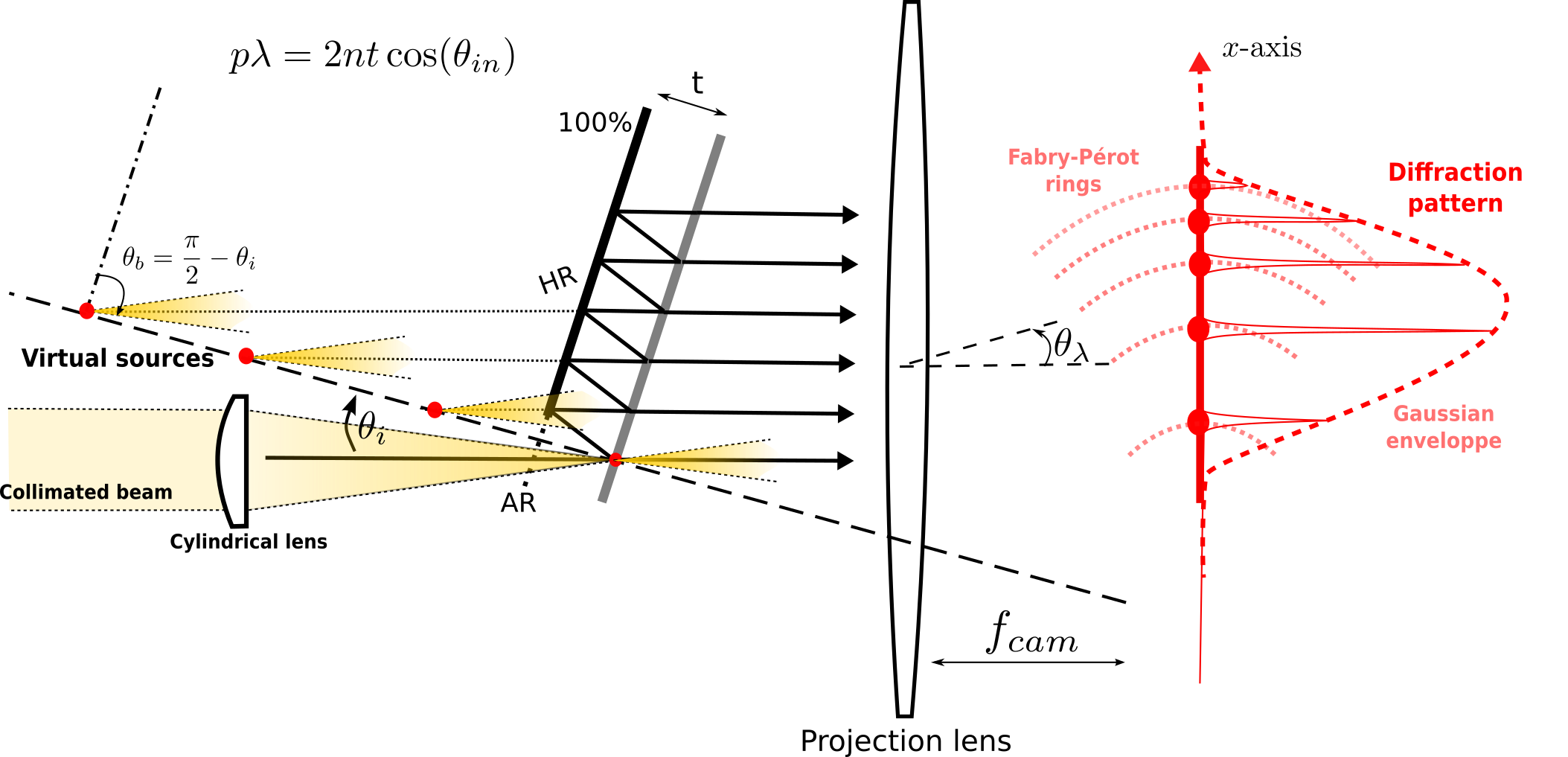}
  \end{tabular}
  \end{center}
  \caption{Physical principle of Virtually Imaged Phased Array (VIPA).}
%>>>> use \label inside caption to get Fig. number with \ref{}
  { \label{fig:vipa_fp_gauss} 
}
\end{figure}

The VIPA (Virtually Imaged Phased Array) is a small optical component which has the physical structure of a Fabry-P\'erot, but is illuminated to behave as an angular disperser, in a way which has some analogy with Lummer-Gehrcke interferometer \cite{Born13}. The VIPA consists of two parallel plane mirrors, the entry side (``back side`` or HR) being almost 100\% reflective, the exit side being highly but partially reflective (PR) $>95$\%. Contrarily to a Fabry-P\'erot, the VIPA is illuminated with a cylindrical lens, whose light is line-focused in a small AR-coated area (with a typical height of a few tens of microns) on the entry reflecting plate, which is the most critical part of the optical alignment. After being injected in the VIPA, light experiences multiple reflections between the two reflective sides, where each reflection can be viewed as a new virtual source \cite{Shirasaki96} with an apodized, exponentialy decreasing, intensity. At the exit of the VIPA, the beams diverge and interfere, in a similar manner to a diffraction grating, such that different wavelengths form constructive interferences in different angular directions, which is schematically represented in Fig \ref{fig:vipa_fp_gauss}.

\subsection{VIPA as a Fabry-P\'erot interferometer}
Assuming an incident gaussian beam of waist $W_0$, the resulting interference pattern is multiplied by a gaussian envelope of waist $W_{env}=\frac{f_{cam}}{f_{cyl}}W_0$. The overall intensity pattern can be described by the function Eq.(\ref{eq:vipa_intensity}), which is consistent with a complete Fourier formalism treatment \cite{Xiao04,Gauthier12}.

\begin{equation}
\label{eq:vipa_intensity}
I(x)=\underbrace{I_0(\lambda)}_{\substack{\text{Normalization} \\ \text{factor}}}\times\underbrace{\exp\left(-2\left(\frac{x}{W_{env}}\right)^2\right)}_{\substack{\text{Gaussian} \\ \text{envelope}}}\times\underbrace{\frac{1}{1+\left(\frac{2}{\pi}F\right)\sin^2\left(\frac{k}{2}\delta\right)}}_{\substack{\text{Fabry-Pérot} \\ \text{interference} \\ \text{function}}}
\end{equation}

with $x$ the 1-D co-ordinate in the detector plane, $F$ the Finesse of the Fabry-P\'erot, $k=\frac{2\pi}{\lambda}$ the wavelength vector, and $\delta=2nt\cos(\theta_i)$ the optical path difference (OPD). The angle in the argument of the OPD is the angle inside the VIPA, as in the well-known Fabry-P\'erot formula\cite{Born13}.
\newline
\newline
As in a Fabry-P\'erot interferometer, each diffraction peak corresponds to an integer value of diffraction order $p$, defined by Eq.(\ref{eq:diff_order}), which is also linked to the resolution of the interferometer by the well-known relation Eq.(\ref{eq:r_i}), which we will call $R$. In addition to the resolution, each spectral element has to be correctly sampled spatially on the detector, which imposes a condition on the angular dispersion of the VIPA given in Eq.(\ref{eq:r_eff}). A more detailed explanation of angular dispersion in the VIPA is given in Sec.~\ref{sec:dispersion}. For a Lorentzian peak, as for Fabry-P\'erot etalon, we found that a slight oversampling of the peak is a more optimal solution, and we set the number $N_S$ of pixels sampling the Lorentzian peak equals to 2.5 pixels ($N_S=2.5$) in our design. In addition, all instruments defects - aberrations in particular - have to be taken into account in the determination of the final resolution.

\begin{gather}
p=\frac{2nt\cos(\theta_i)}{\lambda}
\label{eq:diff_order} \\
R=p\times F 
\label{eq:r_i}\\
R=f_{cam}\frac{\mathrm{d}\theta}{\mathrm{d}\lambda}\mathrm{d}\lambda
\label{eq:r_eff}
\end{gather}

with $t$ the thickness of the VIPA, $n$ the refractive index of the gap into the VIPA, $\theta_i$ the angle of the considered beam \textit{inside the VIPA}, $f_{cam}$ the focal of the camera lens and $\mathrm{d}\lambda=\frac{\lambda}{R_i}$ the smallest resolved spectral element. 

\subsection{Free Spectral Range and echelle spectrometer design}
As in every blaze grating or Fabry-P\'erot, several wavelengths will match the constructive interference condition, resulting in a superposition of several frequencies for one angular direction. In frequency domain, these radiations are regularly spaced by the Free Spectral Range $\Delta\nu=\frac{c}{p\lambda}$. These wavelengths can however be separated by a second disperser, called "cross-disperser", to form the classical scheme of an \textbf{echelle spectrometer}. However, the FSR of the VIPA is relatively small (generally of the order of few hundreds of GHz, or equivalently of few nanometers) given the limited value of Finesse, which constraints the resolution of the Fabry-P\'erot, and thus the thickness and the FSR. This small FSR implies that a large cross-dispersion will be needed, which is provided by a blaze grating as cross-disperser.  

\subsection{VIPA as a virtual diffraction grating}
\label{sec:dispersion}
%As we mentionned, considering each reflection on the back-side as a virtual sources, the VIPA can be seen as a virtual diffraction grating, whose profile is apodized by a exponential decaying intensity function, depending itself on the reflection coefficient and the quality (parallelism, roughness, etc.) of the two plates i.e. finally on the \textbf{Finesse} of the Fabry-Pérot of the interferometer. Thus, as each peaks of the diffraction pattern is related to the magnitude squared of the Fourier transform of the amplitude of the field in the VIPA, these peaks converge to a lorentzian shape for a great number of reflections. These two last intuitive considerations are thus consistent with the Fabry-P\'erot structure of the VIPA.\newline \newline
%Furthermore, this intuitive ``virtual grating`` viewpoint provides a better understanding of the main interesting property of the VIPA, namely its very \textbf{high angular dispersion}. 
Recalling that the angular dispersion of a blaze grating in Littrow configuration is written $\frac{\mathrm{d}\theta}{\mathrm{d}\lambda}=\frac{2\tan(\theta_B)}{\lambda}$, the analogy between a blazed grating and the VIPA viewed as a virtual grating provides relevant order of magnitudes :
\begin{itemize}
\item the equivalent blaze angle of the VIPA is written $\theta_B=\frac{\pi}{2}-\theta_i$, so its angular dispersion is $\propto tan(\frac{\pi}{2}-\theta_i)=\cot(\theta_i)$. With $\theta_i=4^\circ$ , $\cot(4^\circ)\approx 14$. We note that in practice, $\theta_i$ has been commonly used with smaller value as reported in ([weiner, etc.] and that the angular dispersion can be further increased with the index of the gap inside the VIPA.
\item for a blaze grating, angular dispersion is $\propto \tan(\theta_B)$. With $76^\circ$ for a R4 grating, and $63.5^\circ$ for a R2, we obtain $\tan(76^\circ)\approx 4$, and $\tan(63.5^\circ)\approx 2$.
\end{itemize}
We thus see that an increase of at least a factor 3 or 4 in angular dispersion can be expected with the VIPA. As the effective resolution of the VIPA depends of its \textit{Finesse}, very high spectral resolution can be effectively and easily achieved, with a relatively small complexity, the Finesse depending mainly on the quality of the coating, of the polishing and parallelism of the etalon, and can reach nowadays high value at affordable costs ($F=70\; \text{to } 110$ typically). A more detailed treatment of the angular dispersion of the VIPA is given in the next paragraph.

\subsubsection{First order derivation of angular dispersion}
%version 1
%By considering, in a first approach, that $n=1$ (``air-gap`` case), and by differentiating the expression of the OPD , a first order calculation gives an approximated relation of VIPA's angular dispersion, restricted for small angles, given in Eq.(\ref{eq:disp_air}). Considering now that the gap is filled with a material of index $n$, and given the Snell's relation $n\sin(\theta_i)\approx n\theta_i=\theta $, we obtain the relation given in Eq.(\ref{eq:disp_n}).
%
%\begin{align}
%\frac{\mathrm{d}\theta_{\lambda}}{\mathrm{d}\lambda}=-\frac{1}{\lambda_0(\theta_i+\theta_{\lambda})} \label{eq:disp_air} \\
%\frac{\mathrm{d}\theta_{\lambda}}{\mathrm{d}\lambda}=-\frac{n^2}{\lambda_0(\theta_i+\theta_{\lambda})} \label{eq:disp_n}
%\end{align}
%
%, where we separate the angle contribution corresponding to the mechanical tilt of the VIPA $\theta_i$, which is most of the time the major contribution, and the angle $\theta_{\lambda}$ associated with the height co-ordinate along the detector. This two angles $\theta_i$ and $\theta_{\lambda}$ are represented on Fig. \ref{fig:vipa_fp_gauss}. Once again, the very elementary relations provided here are consistent with a complete Fourier formalism, in paraxial approximation, detailed in [ ], which angular dispersion is described by Eq.(\ref{eq:fourier_form}).

Using Fresnel diffraction analysis \cite{Xiao04} with gaussian beams, the angular dispersion of the VIPA can be expressed with Eq.\ref{eq:fourier_form} in paraxial approximation.

\begin{equation}
\label{eq:fourier_form}
\frac{\mathrm{d}\theta_{\lambda}}{\mathrm{d}\lambda}=-\frac{1}{\lambda_0 \left(\frac{\sin(2\theta_i)}{2\left[n^2-\sin^2(\theta_i)\right]}+\frac{\theta_{\lambda}}{n^2}\right)}\stackrel{\text{for } n=1}{=}-\frac{1}{\lambda_0\left(\tan(\theta_i)+\theta_{\lambda}\right)}
\end{equation}
where we separate the angle contribution corresponding to the mechanical tilt of the VIPA $\theta_i$, which is most of the time the major contribution, and the angle $\theta_{\lambda}$ associated with the height co-ordinate along the detector. This two angles $\theta_i$ and $\theta_{\lambda}$ are represented on Fig \ref{fig:vipa_fp_gauss}.
\newline
\newline
At this point, we can also note that the dispersion varies with $\theta_{\lambda}$ (angular dispersion decreases when $\theta_{\lambda}$ increases). We take this fact into account in our design by over-sampling the spectrum at the center of the detector : this ensures that the spectrum on the edge on the detector is correctly sampled although the angular dispersion decreases. For greater tilt angle ($\theta_i>10^\circ$), linear dispersion varies less with $\theta_{lambda}$, but at the cost of a smaller angular dispersion.
\newline
\newline
These relations provide only first-order calculation, which are however useful to a first dimensioning of the instrument. In our design, the OPD was directly computed on Python for the physical optics study, and on the ray-tracing software LASSO for the optical design (LASSO is a optical design software entirely developed by Patrick Rabou, expert optical engineer at IPAG), both being in agreement with the above-mentioned relations. 

\subsubsection{Free-Angular Range and relative size of the envelopppe}\label{sec:FAR}
\begin{figure} [ht]
  \begin{center}
  \begin{tabular}{cc} %% tabular useful for creating an array of images 
  \includegraphics[height=5cm]{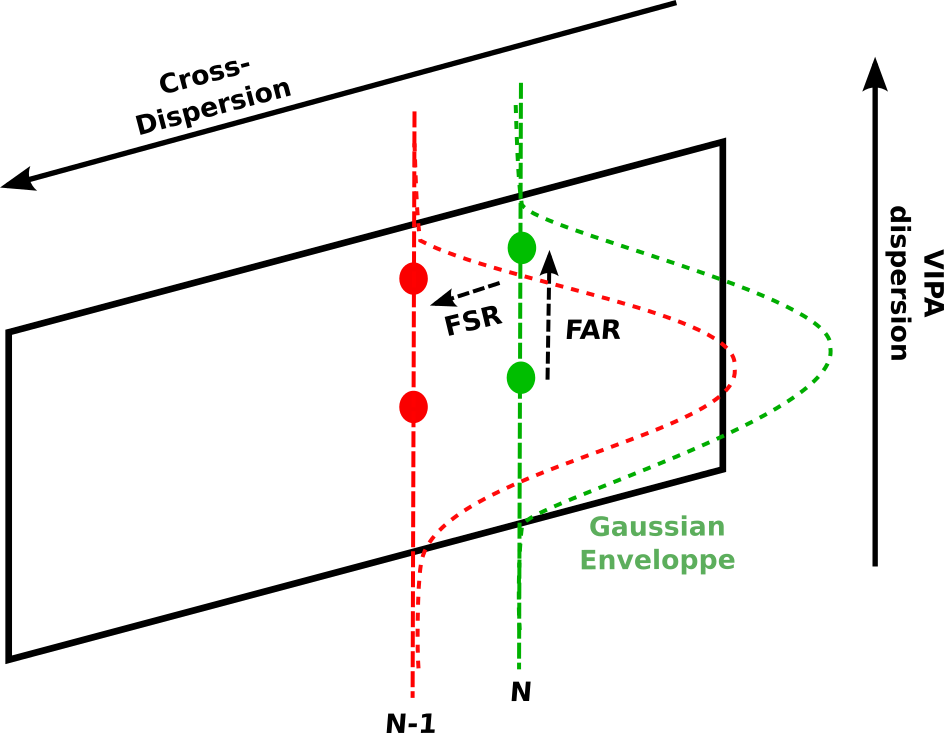} & \includegraphics[height=5cm]{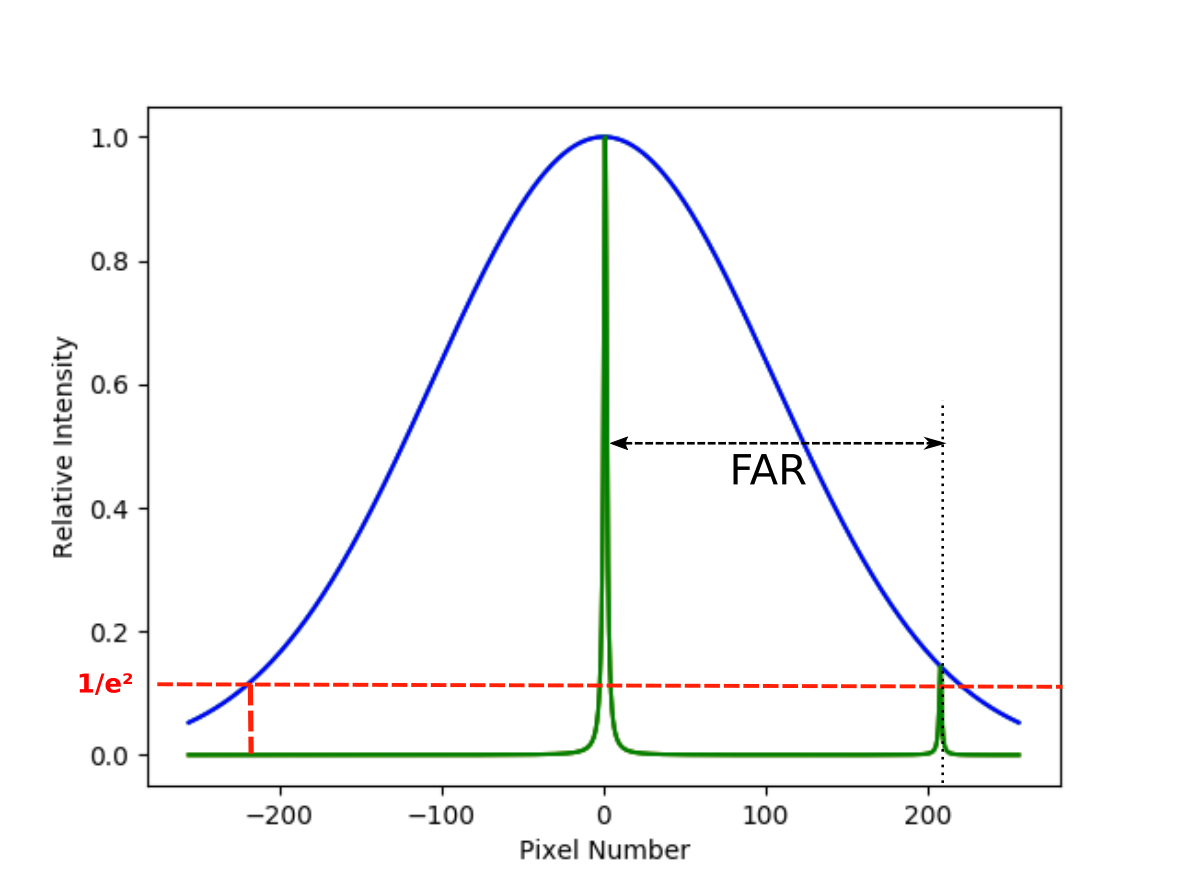}
  \end{tabular}
  \end{center}
  \caption{Illustration of the notions of Free Spectral Range (FSR) and Free Angular Range (FAR) in a design of a cross-dispersed VIPA spectrometer.} 
%>>>> use \label inside caption to get Fig. number with \ref{}
  { \label{fig:FAR_E2} 
}
\end{figure}

From Fig \ref{fig:vipa_fp_gauss}., it can be seen that for one wavelength, several diffraction peaks are however visible on the detector plane, at the position of Fabry-P\'erot rings, which corresponds to the diffraction orders of the VIPA. The different diffraction orders are separated by a variable angular range, called the \textbf{Free Angular Range}, which is directly related to the interference order $p$. As the energy associated with one wavelength will be spread over different peaks, this FAR has to be large compared to the width of the gaussian envelope, in order to concentrate the light in one main peak and to render negligible the contribution of the other ones \cite{Cheng17}. 
\newline
\newline
It has to be noted that once the number of pixels sampling the main diffraction peak has been chosen (the number $N_S$, above mentioned), the FAR expressed in \textit{pixel} units is limited by the finesse $F$ of the VIPA, whose typical value or order of magnitude can be estimated by the relation $\text{FAR}_{pix}=F\times N_S$. Typically, for $F=100$ and $N_S=2.5$, we obtain $\text{FAR}_{pix}=250$, which has to be of the order of the waist of the gaussian envelope, as it is visible on Fig \ref{fig:FAR_E2}. To cover a large number of pixels along $x$-coordinate, it is thus necessary to maximize the Finesse of the VIPA.
\newline
\newline
Analytical description of the FAR are described in literature \cite{Metz14}, and numerical solution were developed under Python and LASSO to take into account carefully this parameter. 

\subsection{Conclusion on the general design}
For a Fabry-P\'erot $R$ can usually reach very high value, and for a VIPA it has been shown in Sec.~\ref{sec:dispersion} that the angular dispersion can be greater by a factor 3 or 4 compared to R4 grating. At a general level, it is of interest to choose high optical index $n$ to maximize the angular dispersion and $F$ to maximize the FAR. However, to increase further the thickness $t$ is generally not interesting, as it will reduced the FSR, needing a greater cross-dispersion.

%schéma simu python avec légende en angle ! Ok
%importance optimiser taille relative enveloppe / FAR : Ok 
%interet d'augmenter (donc diminuer e et augmenter n) et F + : Nok 
  
%At this stage, we thus introduce two quantities related to the VIPA and which are dimensioning in the echelle-spectrometer design :
%
%\begin{itemize}
%\item the \textbf{Free Spectral Range} :
%\item the \textbf{Free angular range} : 
%\end{itemize} 

\section{Optical design of a R=80\,000 infrared echelle-spectrometer}
\subsection{General specifications}
The goal of HDC-Vipa is to demonstrate the working-principle of the VIPA spectrometer for astronomy, in particular to be coupled to extreme Adaptive Optics systems like Sphere or ScExAO. This can be summarized by the following requirements :

\begin{itemize}
\item \textbf{High-Resolution :} the chosen resolution was $R=80\;000$, but it has to be noted that the VIPA has mainly been used for  much larger resolution, from $R=100\,000$ to even $R=1\,000\,000$
\item \textbf{Infrared, cryogenic :} the instrument has to cover $H$ and $K$ bands, and has to be cooled to cryogenic temperature (80\,K)
\item \textbf{Compact and versatile :} the overall instrument has to be easily transportable and adaptable to different other instruments, in particular AO systems and coronagraphs. In our design, we used standard single-mode fiber (SMF), which limit the integrated \textit{\'etendue} to $\lambda^2$
\item \textbf{Multi-channels :} each spectral band has to accept at least two channel i.e. two SMFs.
\end{itemize}

These requirements lead to the following specifications of the HDC-Vipa spectrograph, given in Tab.\ref{tab:gen_spec}.

\begin{table}[ht]
\caption{General Specifications of the spectrometer} 
\label{tab:gen_spec}
\begin{center}       
\begin{tabular}{|c|c|c|} %% this creates columns
%% |l|l| to left justify each column entry
%% |c|c| to center each column entry
%% use of \rule[]{}{} below opens up each row
\hline
\rule[-1ex]{0pt}{3.5ex} \textbf{Channel} & \textbf{H-Band} & \textbf{$K$ band} \\
\hline
\rule[-1ex]{0pt}{3.5ex}  Effective resolution $\lambda/d\lambda$ & \multicolumn{2}{c|}{80\,000}   \\
\hline
\rule[-1ex]{0pt}{3.5ex}  Working temperature &  \multicolumn{2}{c|}{77\,K}  \\
\hline
\rule[-1ex]{0pt}{3.5ex}  Optics platform size & \multicolumn{2}{c|}{400 $\times$ 260\,mm}  \\
\hline
\rule[-1ex]{0pt}{3.5ex}  Vacuum vessel, external dimensions & \multicolumn{2}{c|}{400 $\times$ 500 $\times$ 800\,mm}  \\
\hline
\rule[-1ex]{0pt}{3.5ex}  Spectral domain [nm] & [1500; 1750] & [2000,2400]  \\
\hline
\rule[-1ex]{0pt}{3.5ex}  \textbf{Detector} & \multicolumn{2}{c|}{H2RG, engineering model}   \\
\hline
\rule[-1ex]{0pt}{3.5ex}  Pixel pitch & \multicolumn{2}{c|}{$18\mu m$}   \\
\hline
\rule[-1ex]{0pt}{3.5ex}  Max. area used on detector  & \multicolumn{2}{c|}{512 $\times$ 2048 pixels}  \\
\hline
\rule[-1ex]{0pt}{3.5ex}  Input Fibers & 2 SMFs & 2 SMFs   \\
\hline

\end{tabular}
\end{center}
\end{table} 

The detector is an engineering model of Teledyne H2RG, provided by the Universit\'e de Montr\'eal, which will be particularly useful for the proof of concept. The detector comprises however a significant number of bad pixels which limits its use on sky, but a large portion of these bad pixels are grouped in a half of the detector. They define a usable area of 512 $\times$ 2048 pixels, which corresponds, in height, to the limit imposes by the VIPA and the FAR (see Section \ref{sec:FAR} ), so that it does not impact our design.

%\rule[-1ex]{0pt}{3.5ex}  \textbf{Type} & NUFERN 1550-HP-80 & Thorlabs SM2000  \\
%\hline
%\rule[-1ex]{0pt}{3.5ex}  \textbf{Material} & \multicolumn{2}{c|}{Silica}  \\
%\hline
%\rule[-1ex]{0pt}{3.5ex}  \textbf{Mode-field diameter} & $10.5\mu m$ & $14.5\mu m$  \\
%\hline
%\rule[-1ex]{0pt}{3.5ex}  \textbf{Cladding} & $80 \mu m$ & 2 SMFs, silica  \\
%\hline

\subsection{Design procedure and optimization of the VIPA spectrometer}
As a proof of concept, the design has been optimized to existing off-the-shelf components, excepted for the lens and the two VIPAs which could be customized at a very affordable price. For the VIPA, the goal was to minimize the thickness $t$ while keeping a good injection quality and to obtain a high Finesse (typically, $F=100$) in a material which could endure cryogenic temperature (silica with molecular adhesion). \newline 

The stronger constraints on the design were imposed by the fiber separation, which should be as small as possible but was limited by the cladding diameter (see Tab.\ref{tab:fib+vipa}), and the trade-off between dispersive power of the cross-disperser and efficiency. However, this constraint on fiber separation is not a fundamental limit and could be overcame with carefully designed components, such as multi-core fibers or integrated optics components. Such components are also in use in several astronomical instruments, the goal being to minimize the size of the mode-field diameter and to group a maximum number of channels on each order on the detector. Since off-the-shelf fluoride glass fibers do not comply with our requirements, it imposes the choice of fiber thorlabs SMF2000 which is clearly not optimal in terms of attenuation and could be further optimized in an upgrade of the instrument.
\newline  

%\begin{table}[ht]
%\caption{VIPA and fibers characteristics} 
%\label{tab:fib+vipa}
%\begin{center}       
%\begin{tabular}{|c|c|c||c|c|p{3cm}|} %% this creates columns
%% |l|l| to left justify each column entry
%% |c|c| to center each column entry
%% use of \rule[]{}{} below opens up each row
%\hline
%\rule[-1ex]{0pt}{3.5ex}  \textbf{VIPA} & $H$ band & $K$ band & \textbf{Fibers} & $H$ band & $K$ band \\
%\hline
%\rule[-1ex]{0pt}{3.5ex}  \textbf{Domain} &$1.5-1.75\mu$m & $2.0-2.4\mu$m& \textbf{Domain} & $1.5-1.75$\,$\mu$m & $2.0-2.4$\,$\mu$m  \\
%\hline
%\rule[-1ex]{0pt}{3.5ex}  \textbf{Material} & \multicolumn{2}{c||}{Fused silica} & \textbf{Material} & \multicolumn{2}{c|}{Silica} \\
%\hline
%\rule[-1ex]{0pt}{3.5ex}  \textbf{Finesse} & $>110$ & $>105$ &  \textbf{Reference} & Nufern 1550-HP-80 & Thorlabs SM2000  \\
%\hline
%\rule[-1ex]{0pt}{3.5ex}  \textbf{HR Coating} & $>99.8\%$ & $>99.8\%$ & \textbf{NA} & $0.13$ & $0.12$ \\
%\hline
%\rule[-1ex]{0pt}{3.5ex}  \textbf{PR Coating}  & $96\%$ & $95\%$ & \textbf{MFD} & 10$\mu$m @1.5\,$\mu$m & 13\,$\mu m @2\mu m$  \\
%\hline
%\rule[-1ex]{0pt}{3.5ex}  \textbf{Dimensions} & \multicolumn{2}{c||}{30 $\times$ 24\'mm} & \textbf{Cladding diam.} & $80\mu m$ & $125\mu m$  \\
%\hline
%\rule[-1ex]{0pt}{3.5ex}  \textbf{Entry length} & 0.25mm & 0.3mm & %\textbf{Attenuation} & 0.5dB/km@$1.5\mu$m & $20dB/km@2\mu$m \\
%\hline
%\rule[-1ex]{0pt}{3.5ex}  \textbf{Tilt Angle} & \multicolumn{2}{|c||}{$\theta_i=4.55^\circ$} & & & $250dB/km@2.3\mu$m \\
%\hline
%
%\end{tabular}
%\end{center}
%\end{table} 

\begin{table}[ht]
\caption{VIPA and fibers characteristics} 
\label{tab:fib+vipa}
\begin{center}       
\begin{tabular}{|c|c|c|} %% this creates columns
%% |l|l| to left justify each column entry
%% |c|c| to center each column entry
%% use of \rule[]{}{} below opens up each row
%\hline
 \multicolumn{3}{c}{\textbf{VIPA} }\\ \hline
 & $H$ band & $K$ band  \\ \hline
\textbf{Domain} & $1.5-1.75\mu$m & $2.0-2.4\mu$m\\ \hline
\textbf{Material} & \multicolumn{2}{c|}{Fused silica} \\ \hline
\textbf{Finesse} & $>110$ & $>105$  \\ \hline
\textbf{HR Coating} & $>99.8\%$ & $>99.8\%$  \\ \hline
\textbf{PR Coating}  & $96\%$ & $95\%$ \\\hline
\textbf{Dimensions} & \multicolumn{2}{c|}{30 $\times$ 24\'mm} \\ \hline
\textbf{Entry length} & 0.25mm & 0.3mm\\\hline
\textbf{Tilt Angle} & \multicolumn{2}{c|}{$\theta_i=4.55^\circ$}  \\ \hline

 \multicolumn{3}{c}{\textbf{Fibers} }\\ \hline

& $H$ band & $K$ band \\ \hline
\textbf{Domain} & $1.5-1.75$\,$\mu$m & $2.0-2.4$\,$\mu$m  \\ \hline
\textbf{Material} & \multicolumn{2}{c|}{Silica} \\ \hline
\textbf{Reference} & Nufern 1550-HP-80 & Thorlabs SM2000  \\ \hline
\textbf{NA} & $0.13$ & $0.12$ \\ \hline
\textbf{MFD} & 10$\mu$m @1.5\,$\mu$m & 13\,$\mu m @2\mu m$  \\ \hline
\textbf{Cladding diam.} & $80\mu m$ & $125\mu m$  \\ \hline
\textbf{Attenuation} & 0.5dB/km@$1.5\mu$m & $20dB/km@2\mu$m \\
& & $250dB/km@2.3\mu$m \\ \hline
 
\end{tabular}
\end{center}
\end{table}

For $H$ band, a 300g/mm in order 3, $\theta_B=46^\circ$ grating, with protected silver coating is used as cross-disperser. For $K$ band, the cross-disperser is a 600g/mm in order 1, $\theta_B=54^\circ$ grating. Both are off-the-shelf components. It has to be noted that, considering the small FSR of the VIPA, a large cross-dispersion is needed, which has been an important constraint in our design.
\newline
\newline
Given the different constraints on the design (detectors, fibers, maximal Finesse), trade-off were set for the design. This design hierarchy is described in Tab.\ref{tab:dh}.

\begin{table}[ht]
\caption{Hierarchy of the parameters in the VIPA spectrograph design. } 
\label{tab:dh}
\begin{center}       
\begin{tabular}{|c|c|c|} %% this creates columns
%% |l|l| to left justify each column entry
%% |c|c| to center each column entry
%% use of \rule[]{}{} below opens up each row
\hline
\rule[-1ex]{0pt}{3.5ex} \textbf{Input Parameter} & \textbf{Requirements} & \textbf{Derived Parameter} \\
\hline
\rule[-1ex]{0pt}{3.5ex} Detectors \& Max. Finesse & FAR & VIPA tilt angle \\
\hline
\rule[-1ex]{0pt}{3.5ex} VIPA tilt angle & Dispersion & $f_{cam}$ \\
\hline
\rule[-1ex]{0pt}{3.5ex} $f_{cam}$ &   &   \\
\rule[-1ex]{0pt}{3.5ex} Fiber core \& spacing & Fiber spacing on detector & $f_{col}$ \\
\rule[-1ex]{0pt}{3.5ex} Grating &   &  \\
\hline
\rule[-1ex]{0pt}{3.5ex} $W_0$ & Relative size of gaussian enveloppe vs FAR & $f_{cyl}$ \\
\hline

\end{tabular}
\end{center}
\end{table} 

\subsection{Instrument description}
\label{sec:eff}
\begin{figure} [ht]
  \begin{center}
  \begin{tabular}{c} %% tabular useful for creating an array of images 
  \includegraphics[height=7.5cm]{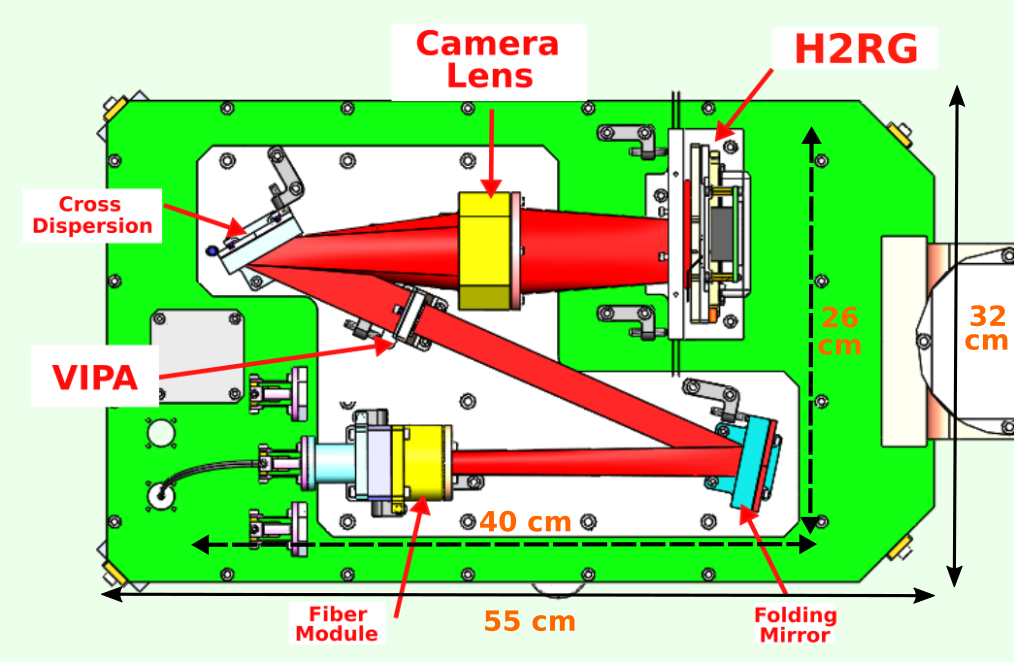}  
  \end{tabular}
  \end{center}
  \caption{Optical platform in $H$ band, viewed from above.} 
%>>>> use \label inside caption to get Fig. number with \ref{}
  { \label{fig:above_leg} 
}
\end{figure}

\begin{figure} [ht]
  \begin{center}
  \begin{tabular}{c} %% tabular useful for creating an array of images 
  \includegraphics[height=10cm]{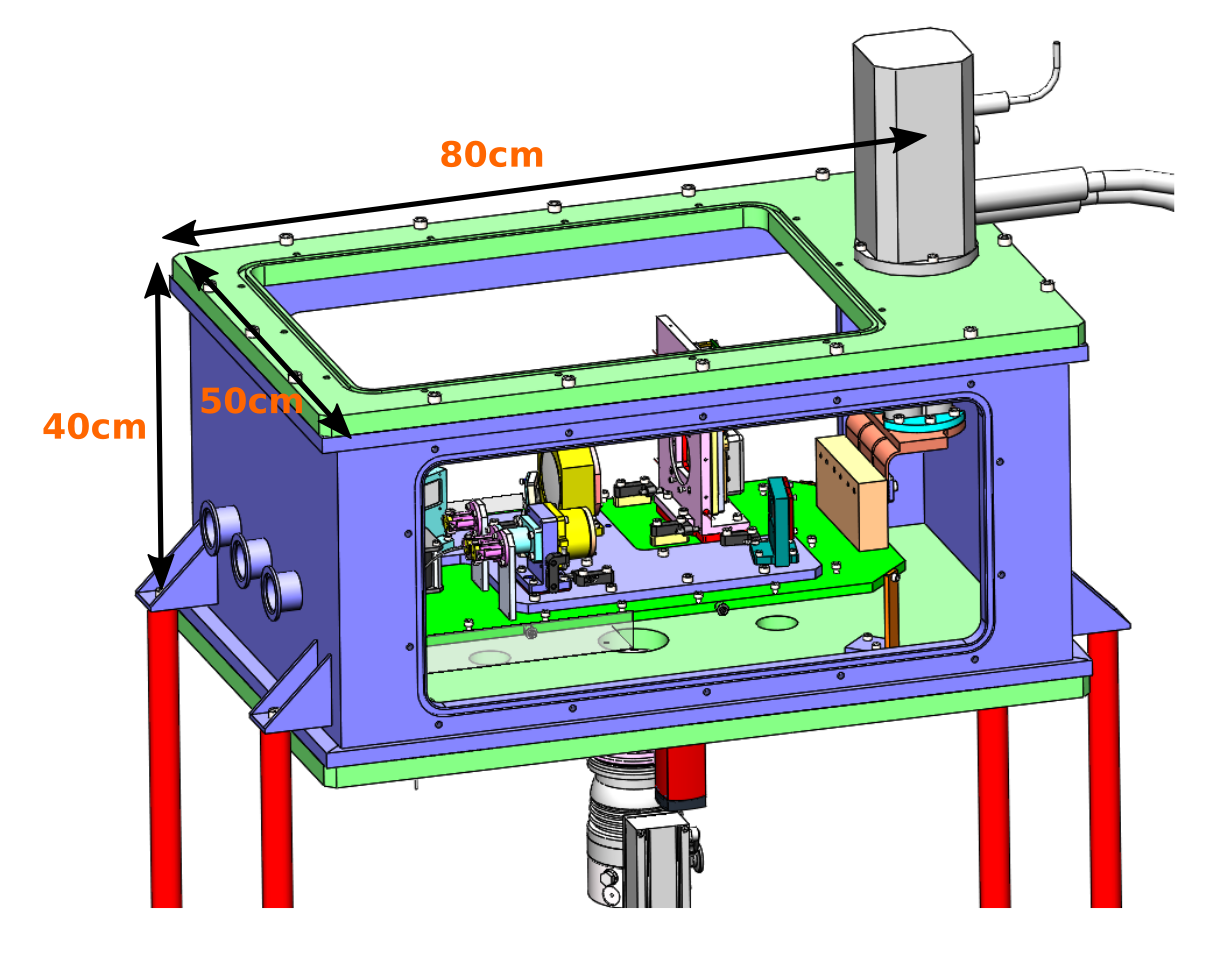}  
  \end{tabular}
  \end{center}
  \caption{Optical platform in $H$ band viewed in the vacuum vessel}
%>>>> use \label inside caption to get Fig. number with \ref{}
  { \label{fig:vacuum+vessel} 
}
\end{figure}

The resulting instrument is a spectrograph placed on a 400 $\times$ 260\,mm platform, which fits within a vacuum vessel whose maximum dimensions are 400 $\times$ 500 $\times$ 800\,mm. For each Band ($H$ and $K$), an optical platform with its own optics optimized for the given spectral band has been built, each optical platform being aligned and tested separately, and then set in the vacuum vessel, where the detector is fixed permanently. A view of the optical platform and the vacuum vessel are given in Fig \ref{fig:above_leg}. and Fig \ref{fig:vacuum+vessel}. For simplicity reasons, the two optical platforms cannot be used simultaneously in this first demonstrator, but this is certainly doable in a future upgrade, for instance in order to project the 2 bands on different parts of a fully operational H2RG (2K $\times$ 2K) detector.

 For the moment, the two optical platforms cannot be used at the same time on the detector, which was imposing several constraints on the optical design, but their simultaneous use has to be planned for an upgrade of the instrument.

\subsubsection{Acquisition camera optics}
\begin{figure} [ht]
  \begin{center}
  \begin{tabular}{cc} %% tabular useful for creating an array of images 
  \includegraphics[height=6cm]{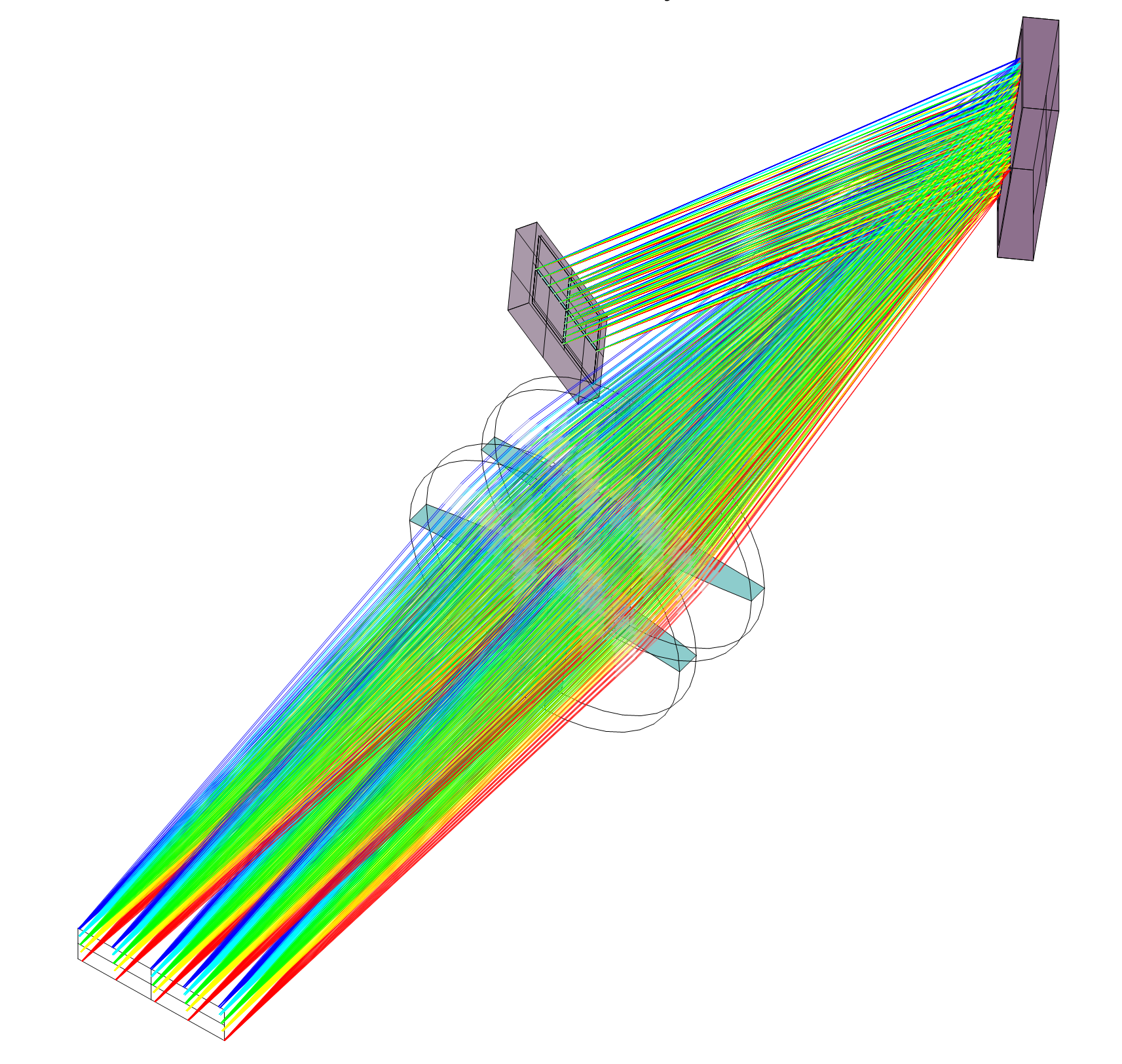} &
  \includegraphics[height=6cm]{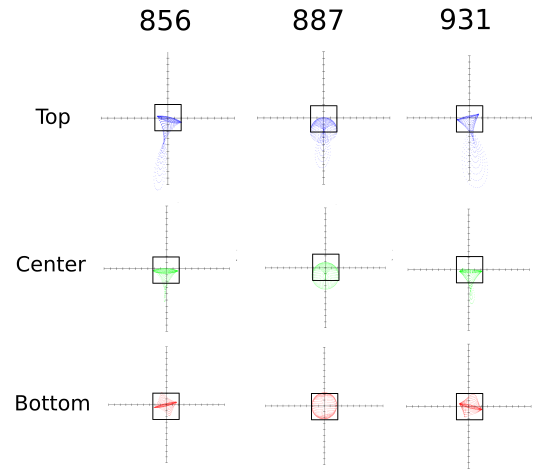} 
  \end{tabular}
  \end{center}
  \caption{VIPA, cross-disperser and camera, viewed on the ray-tracing software LASSO ; Spot-diagram for $H$ band : in the diagram, each black square represents the size of a pixel of the H2RG. } 
%>>>> use \label inside caption to get Fig. number with \ref{}
  { \label{fig:camera} 
}
\end{figure}

The acquisition camera lens is a doublet of focal $f_{cam}=150mm$ for $H$ and $K$-band designs, and is made of S-TIH14, which is visible on Fig \ref{fig:camera}. In comparison to the best optical design, a slight curvature has been added to plane surfaces of the lenses to avoid ghost reflections with the surface of the detector. 

\subsubsection{Overall throughput}
Considering all the optical elements of the spectrograph, from fiber attenuation to detector QE, the estimated throughput is $\eta=0.4$ for $H$ band and $\eta=0.3$ for $K$ band. In $K$ band, several elements could be easily optimized, in particular fiber attenuation, with the use of ZFG fibers, and VIPA vignetting, with a bigger width of our VIPA in $K$ band. The estimated throughput is given in Tab.\ref{tab:eff}. However, it is important to note that in essence, given its structure of Fabry-P\'erot interferometer, the VIPA has an optimal throughput,  which is no longer limited by the HR coating ($>0.998$ in our case), but by the quality of injection in the VIPA.

\begin{table}[ht]
\caption{Estimated throughput of the end-to-end spectrograph. Important losses are due to the grating, a point which could be optimized with VPH grating for example. In $K$ band, due to extra-losses in fiber and at the VIPA injection the overall throughput is smaller, but better performances could be expected with dedicated, optimised components.  } 
\label{tab:eff}
\begin{center}       
\begin{tabular}{|c|c|c|} %% this creates two columns
%% |l|l| to left justify each column entry
%% |c|c| to center each column entry
%% use of \rule[]{}{} below opens up each row
\hline
\rule[-1ex]{0pt}{3.5ex}  Item & \textbf{$H$ band} & \textbf{$K$ band}  \\
\hline
\rule[-1ex]{0pt}{3.5ex}  5\,m Fiber & 100\% & 84\% \\
\hline
\rule[-1ex]{0pt}{3.5ex}  Relay Optics & 80\% & 80\% \\
\hline
\rule[-1ex]{0pt}{3.5ex}  VIPA & 100\% & 80\% \\
\hline 
\rule[-1ex]{0pt}{3.5ex}  Grating & 70\% & 80\%  \\
\hline
\hline
\rule[-1ex]{0pt}{3.5ex}  Spectro itself, incl. fibers & 56\% & 43\%  \\
\hline
\hline
\rule[-1ex]{0pt}{3.5ex}  H2RG QE & \multicolumn{2}{c|}{70\%}   \\
\hline
\rule[-1ex]{0pt}{3.5ex}  Total, incl. QE & 40\% & 30\%  \\
\hline
\end{tabular}
\end{center}
\end{table} 

%\subsection{Limitation by the angle of incidence $\theta_i$}
%-> schéma plus calcul angle limite

%\subsubsection{Cross-dispersion and fibers : trade-off}
%-> compromis avec composant dispo
%-> tend à favoriser fiber à très petit coeur

\subsection{Cryogenic study}
The cryogenic study has been an important step in the design, given the sensitivity of the injection of light into the VIPA. The approach adopted for the cryogenic study was :

\begin{itemize}
\item \textbf{1) Optical design @77\,K :} the optical design on ray-tracing software LASSO has been done at the nominal working temperature 77\,K and for vacuum.
\item \textbf{2) Mechanical design @293\,K : } taking into account the integrated variation of CTE over the temperature range of both the optics (S−TIH14, CaF2 and fused silica) and the optical bench (Al-6061), and the variation of optical index with temperature, the mechanical design and radii of curvature of the optics were defined at 293\,K. The data for S-TIH14 were calculated from experimental data on S-TIH1\cite{Manuel13}.
\item \textbf{3) Verification @77\,K :} finally, a temperature gradient and vacuum is applied on the whole final design to verify the image quality and the alignment on LASSO.   
\end{itemize} 

Experimentally, the different adjustments will be done at T=293\,K, with different iterations to set-up the optics.

\subsection{Optical Design : Conclusion}
From the previous general guidelines, that allow a wide diversity of potentiel VIPA design, we show here that a choice of VIPA design parameter can be selected to satifsfy the need of high resolution spectroscopy in NIR optimized for exoplanet atmosphere molecular analysis. Such parameter are very well compatible with current manufacturing capabilities at low cost.

\section{Further steps and outlook}
\subsection{Planning}
For the $H$-band platform, integration tests are planned for the Quarter 4 of 2018, and the tests up to the end of Quarter 1 of 2019. Calibration procedures, including calibration lamp and Fabry-P\'erot, have to be set in the meantime. In parallel to this experimental work, the mechanical design study of $K$ band will be finished during the summer 2018. The purpose of this development is to conclude on the possibility to couple the instrument with Adaptive Optics in the context of HDC for Quarter 2 of 2019.

\subsection{Optimization }
Several steps could be envisioned as part of the proof-of-concept to optimize further the instrument :

\begin{itemize}
\item  reduce the spacing between the input channels of the spectrograph and increase their numbers. This point has been identified as an important constraint during the design, and could be addressed with dense multi-core fiber, developed currently in the context of high-dispersion coronography \cite{Haffert18}, or integrated optics components, developed in astronomical interferometry;
\item  optimize the throughput, as mentionned in Sec~\ref{sec:eff};
\item  merge the two platforms in order to use simultaneously $H$ and $K$ bands on the H2RG;
\item study the coupling between AO system and SMF fiber, and the possibility and limitations to pave the focal plane prior to the spectrograph.
\end{itemize}

\section{Conclusion}
We present the design of a $R=80\,000$, cryogenic infrared spectrograph developed at IPAG. The concept of the instrument is based on the use of VIPA, a kind of Fabry-P\'erot interferometer used as a virtual grating, which enables to reach high resolution in a compact design. The spectrograph covers $H$ and $K$ bands, with two different optical platforms, which can not be used simultaneously in a first time, each platform having 2 input SMFs. Laboratory tests are planned from Q4 2018 to Q2 2019, in order to characterize the instrument demonstrate the feasibility of a coupling with adaptive optics system as SPHERE in the context of high-dispersion coronography. Finally, this approach results in an instrument that is simple to manufacture and whose design is adaptable to different needs in terms of spectral range, illuminated zone on the detector, or resolution, noting in particular the capability of the VIPA to reach high or even very high ($R>100\,000$) spectral resolution.
%natural tendency of the VIPA to reach high-resolution

\acknowledgments % equivalent to \section*{ACKNOWLEDGMENTS}       

This research was supported by the LabEx FOCUS ANR-11-LABX-0013. The authors thank the LabEx FOCUS for its help and funding. \\
The authors thank the Universit\'e of Montr\'eal for providing the detector and its collaboration.\\
Finally, the first author thanks Alexandre Jeanneau for his support and the fruitful discussions with him during the paper redaction. 

% References
\bibliography{report} % bibliography data in report.bib

\begin{thebibliography}{10}

\bibitem{Riaud07}
Riaud, P. and Schneider, J., ``Improving earth-like planets' detection with an
  elt: the differential radial velocity experiment,'' {\em AA}~{\bf 469},
  355--361 (2007).

\bibitem{Lovis16}
{Lovis}, C. and al, ``Atmospheric characterization of proxima b by coupling the
  sphere high-contrast imager to the espresso spectrograph,'' {\em ArXiv
  e-prints} ,  arXiv:1609.03082 (Sept. 2016).

\bibitem{Snellen2013}
Snellen, I. A.~G., de~Kok, R.~J., le~Poole, R., Brogi, M., and Birkby, J.,
  ``Finding extraterrestrial life using ground-based high-dispersion
  spectroscopy,'' {\em The Astrophysical Journal}~{\bf 764}(2),  182 (2013).

\bibitem{Kasper2016}
{Wagner}, K., {Apai}, D., {Kasper}, M., {Kratter}, K., {McClure}, M.,
  {Robberto}, M., and {Beuzit}, J.-L., ``{Direct imaging discovery of a Jovian
  exoplanet within a triple-star system},'' {\em Science}~{\bf 353},  673--678
  (Aug. 2016).

\bibitem{Wang17}
Wang, J., Mawet, D., Ruane, G., Hu, R., and Benneke, B., ``Observing exoplanets
  with high dispersion coronagraphy. i. the scientific potential of current and
  next-generation large ground and space telescopes,'' {\em The Astronomical
  Journal}~{\bf 153}(4),  183 (2017).

\bibitem{mawet17}
{Mawet}, D., {Ruane}, G., {Xuan}, W., {Echeverri}, D., {Klimovich}, N.,
  {Randolph}, M., {Fucik}, J., {Wallace}, J.~K., {Wang}, J., {Vasisht}, G.,
  {Dekany}, R., {Mennesson}, B., {Choquet}, E., {Delorme}, J.~R., and
  {Serabyn}, E., ``Observing exoplanets with high-dispersion coronagraphy. ii.
  demonstration of an active single-mode fiber injection unit,'' {\em APJ}~{\bf
  838},  92 (Apr. 2017).

\bibitem{Jovanovic16}
Jovanovic, N., Schwab, C., Cvetojevic, N., Guyon, O., and Martinache, F.,
  ``Enhancing stellar spectroscopy with extreme adaptive optics and
  photonics,'' {\em Publications of the Astronomical Society of the
  Pacific}~{\bf 128}(970),  121001 (2016).

\bibitem{Lexi16}
{Haffert}, S.~Y., {Wilby}, M.~J., {Keller}, C.~U., and {Snellen}, I.~A.~G.,
  ``The leiden exoplanet instrument (lexi): a high-contrast high-dispersion
  spectrograph,'' in [{\em Ground-based and Airborne Instrumentation for
  Astronomy VI}{\nolinebreak\hspace{0.1em}]},   {\bf 9908},  990867 (Aug.
  2016).

\bibitem{Jacquinot54}
Jacquinot, P., ``The luminosity of spectrometers with prisms, gratings, or
  fabry-perot etalons,'' {\em J. Opt. Soc. Am.}~{\bf 44},  761--765 (Oct 1954).

\bibitem{Schroeder2000}
Schroeder, D.,  [{\em Astronomical Optics}{\nolinebreak\hspace{0.1em}]},
  Electronics \& Electrical, Academic Press (2000).

\bibitem{Shirasaki96}
Shirasaki, M., ``Large angular dispersion by a virtually imaged phased array
  and its application to a wavelength demultiplexer,'' {\em Opt. Lett.}~{\bf
  21},  366--368 (Mar 1996).

\bibitem{Diddams07}
Diddams, S.~A., Hollberg, L., and Mbele, V., ``Molecular fingerprinting with
  the resolved modes of a femtosecond laser frequency comb,'' {\em Nature}~{\bf
  445}(7128),  627 (2007).

\bibitem{Bourdarot17}
Bourdarot, G., Le~Coarer, E., Bonfils, X., Alecian, E., Rabou, P., and Magnard,
  Y., ``Nanovipa: a miniaturized high-resolution echelle spectrometer, for the
  monitoring of young stars from a 6u cubesat,'' {\em CEAS Space Journal}~{\bf
  9}(4),  411--419 (2017).

\bibitem{Born13}
Born, M. and Wolf, E.,  [{\em Principles of Optics: Electromagnetic Theory of
  Propagation, Interference and Diffraction of
  Light}{\nolinebreak\hspace{0.1em}]}, Elsevier Science (2013).

\bibitem{Xiao04}
Xiao, S., Weiner, A.~M., and Lin, C., ``A dispersion law for virtually imaged
  phased-array spectral dispersers based on paraxial wave theory,'' {\em IEEE
  journal of quantum electronics}~{\bf 40}(4),  420--426 (2004).

\bibitem{Gauthier12}
Gauthier, D.~J., ``Comment on “generalized grating equation for virtually
  imaged phased-array spectral dispersers”,'' {\em Applied optics}~{\bf
  51}(34),  8184--8186 (2012).

\bibitem{Rabou}
Software entirley developed~by Patrick~Rabou, e. o. e. a.~I.

\bibitem{Cheng17}
ChengZe~Song, Emilio Sánchez-Ortiga, M. R. F. P.~T., ``Optimisation of a
  single stage vipa spectrometers (conference presentation),'' {\em
  Proc.SPIE}~{\bf 10067},  10067 -- 10067 -- 1 (2017).

\bibitem{Metz14}
Metz, P., Adam, J., Gerken, M., and Jalali, B., ``Compact, transmissive
  two-dimensional spatial disperser design with application in simultaneous
  endoscopic imaging and laser microsurgery,'' {\em Applied optics}~{\bf
  53}(3),  376--382 (2014).

\bibitem{Manuel13}
Manuel A.~Quijada, Douglas B.~Leviton, D. A.~C., ``Cryogenic refractive index
  and coefficient of thermal expansion of s-tih1 glass,'' {\em Proc.SPIE}~{\bf
  8863},  8863 -- 8863 -- 10 (2013).

\bibitem{Haffert18}
S., H., E., P., C., K., and I., S., ``Using multi-core fiber technology to
  couple sphere to high-resolution integral-field spectrographs,'' {\em SPHERE
  Upgrades Workshop}  (2018).

\end{thebibliography}
\bibliographystyle{spiebib} % makes bibtex use spiebib.bst

\end{document}